\def\be{\begin{equation}}
\def\ee{\end{equation}}
\def\bea{\begin{eqnarray}}
\def\eea{\end{eqnarray}}
\def\bq{\begin{quote}}
\def\eq{\end{quote}}

\def\gappeq{\mathrel{\rlap {\raise.5ex\hbox{$>$}}
{\lower.5ex\hbox{$\sim$}}}}

\def\lappeq{\mathrel{\rlap{\raise.5ex\hbox{$<$}}
{\lower.5ex\hbox{$\sim$}}}}

\documentstyle [12pt]{article}

\begin{document}
\thispagestyle{empty}
%\vspace*{-2cm}
\begin{flushright}
{MPI-PhT-96-16} \\
{March 1995} \\
\end{flushright}
\vspace{1cm}
\begin{center}
{\large Ingredients and Equations for making a} \\
\vspace{.2cm}
{\large Magnetic Field in the Early Universe} \\
%\vspace{.2cm}
%{\large }
\end{center}
\vspace{1cm}
\begin{center}
{Sacha Davidson }\\
\vspace{.3cm}
{Max Planck Institut f\"{u}r Physik\\
F\"{o}hringer Ring 6, 80805, M\"{u}nchen, Germany} % \\
%fax: (49) (89) 32267-04 , office: (49) (89) 32354-251\\
%email: sacha@mppmu.mpg.de}
\end{center}
\vspace{1cm}
\hspace{3in}

\begin{abstract}
The ingredients required  to create
 a magnetic field in the early Universe
are identified, and compared with
Sakharov's conditions for baryogenesis. 
It is also shown that  a long range coherent magnetic
field is not generated by the classical
rolling Higgs vacuum expectation value during the 
electroweak phase transition.

\end{abstract}

\newpage

The Universe is observed to have coherent magnetic fields
over a wide variety of scales, extending from
the earth to possibly galaxy clusters \cite{K}. 
These observations are puzzling, because
it is not clear how the large-scale fields are made.

Our galactic disk locally has
a coherent magnetic field of strength $\sim 10^{-6}$ Gauss,
approximately in the direction of the galactic arm \cite{K,Rees}.
Observations of the Faraday rotation of 
radio waves from extragalactic sources by
various spiral galaxies suggest that these galaxies 
today  have aximuthal magnetic
fields of strength $\sim  3 \times 10^{-6}$ Gauss.
 There are also recent data suggesting micro Gauss
fields are common in galaxy clusters. Measuring
magnetic fields at high redshift is difficult, but
it appears that there is a galaxy at $z \sim .4$  with
a microGauss field, and that  the Lyman $\alpha$ forest
may also include clouds with fields of
about this strength \cite{K}.
An extensive review of the observations and astrophysics
of magnetic field preservation and amplification can be found
in \cite{K}.

An aximuthal field is a natural configuration inside
a differentially rotating spheroid, and suggests that
the present magnetic field in spiral galaxies may have
been amplified by the dynamo mechanism \cite{K,P}. 
However, the timescale during which the dynamo grows
the field, and  by  how much the field is magnified,
are unclear, so although one
knows that a   seed field is required on galactic scales, 
its magnitude is not well defined
($\vec{B} \sim 10^{-10}$ ---$ 10^{-24}$ Gauss).
There are various astrophysical \cite{K} and
cosmological \cite{H,V,max,inf,McL} scenarios for
making this seed field.

This paper addresses two distinct issues related
to the generation of magnetic fields.  In the first
section, the symmetries, or properties,
of the early Universe that are broken by the
presence of magnetic fields are identified,
and whether they are required to {\it generate} a
 field is discussed in the light of various
models.
Interestingly enough, these ``broken properties'', which
are $C, CP$, thermal equilibrium and
rotational invariance, are very similar to
Sakharov's required ingredients for baryogenesis.

The second issue addressed by this paper is
whether the rolling higgs vev at the electroweak
phase transition (EPT)  produces  a coherent,
horizon-scale magnetic field. One of
the popular cosmological mechanisms
for generating the seed magnetic field
is due to Vachaspati \cite{V}, who
pointed out that there are Higgs
field contributions to the 
electromagnetic $F^{\mu \nu}$ during
the EPT. This is frequently taken
to mean that horizon-scale magnetic fields of
strength $\sim \alpha T^2$ are created at the
phase transition. In the second
section of this paper, it is shown that the 
rolling Higgs vev is electrically neutral.
This means that  classically, there are no sources
for the magnetic field during the phase
transition, and no long range field is generated.

\section{making a magnetic field}

 Three ingredients are neccessary 
to make a baryon asymmetry: one must have
baryon number violation, $C$ and $CP$ violation, and 
 out of equilibrium dynamics. The
familiar explanation for this is that baryon number is odd
under $C$ and $CP$, and there are no asymmetries
of non-conserved quantities in a thermal bath. 
Reasoning by analogy,  one would like to know   what
ingredients are required to make a magnetic field
in the early Universe? 

 Let us first address  a simpler question: not
what is required to create a magnetic field, but rather
what symmetries are broken, or what conditions
do not apply, in  the presence of
a magnetic field? This gives  
three conditions:
 one needs some kind of out-of-thermal-equilibrium dynamics, 
because in  equilibrium the photon
distribution is thermal, and there are no particle currents to
sustain a ``long range''   field. 
One  needs $C$ and $CP$ violation, because  the $\vec{B}$ field
is odd under both, and the initial state of the
Universe is expected to be even under these symmetries.
And finally, one needs to break the isotropy of the
early Universe, because the  magnetic  field chooses a
direction. So the properties of the early
Universe that are broken by the presence
of a magnetic field ($C$, $CP$, $SO(3)$ and
thermal equilibrium) are very similar to those
broken by the baryon asymmetry \cite{Sak}. 

The interesting question, however, is
not what properties are broken
by the presence of the magnetic field, but
rather  what ingredients
are required to {\it make} it. In
the case of the baryon asymmetry, the symmetries
that are broken by the presence of the particle
excess have to be broken to generate it.  But
for the magnetic field, this is not the case,
because it is a classical field.  
It can
``spontaneously'' develop an
expectation value, if it has an
effective potential of the right shape. So
although many models for the generation
of a primordial seed magnetic field do
contain explicit $C$, $CP$ and $SO(3)$ breaking,
this is in principle not {\it required}.
The analogy between generating the baryon asymmetry
and a magnetic field can clarify this. In baryogenesis,
the desired end result is a $C$, $CP$ and baryon number
violating excess of particles over anti-particles. 
It is possible to generate this asymmetry
using spontaneous $CP$ \cite{A+D} and baryon number \cite{WW}
violation; in other words, the Lagrangian need
not break either of these
symmetries explicitly. However, prior
to the generation
of the perturbative particle asymmetry,  a classical
field develops  a baryon number and/or $CP$ violating
vev. So that from the  point of view of making
the excess of particles over anti-particles,
baryon number and $CP$ violation are required.
The magnetic field, on the other hand,
is a classical object, and can itself
spontaneously break $C$, $CP$ and
$SO(3)$ as it develops. (Note that this
implies, in principle, that one could
use a magnetic field to provide the $CP$ violation
required in baryogenesis models.)
The only requirement for
magnetic field generation is therefore some
out of thermal equilibrium  physics.

 Let us now
consider various models for the generation
of a $\vec{B}$ field, to illustrate  the above
discussion.
One way to generate a primordial magnetic field
is to make an electric current, which
has an associated magnetic field \cite{max}. In this case,
one does need explicit $C$, $CP$ and $SO(3)$ violation,
as well as some kind of out of equilibrium dynamics, because
these are required  to generate an electromagnetic current
in the particles present in the hot early Universe.
This can make it difficult to generate a sufficiently
large seed field, because $CP$ violation
and out of equilibrium physics are hard to come by
in the early Universe. 
 Astrophysical mechanisms \cite{K} for
the generation of the seed magnetic field also
make the field ``via Maxwell's equations'', and
require that all these conditions be broken
explicitly. However, this is not a significant constraint
by the time galaxies are formed, because $C$,
$CP$ and $SO(3)$ are no longer approximately
symmetries of the Universe. 
A  primordial
magnetic field can also be generated by amplifying fluctuations during
the inflationary epoch \cite{inf}. 
 The intuitive picture is that a small 
fluctuation in the electromagnetic field
is inflated into a classical field. 
This is clearly a non-equilibrium
process, but no explicit symmetry breaking is
required. Another mechanism that does
not require explicit $C$, $CP$ and $SO(3)$ violation is
to amplify magnetic field fluctuations using
the dynamo mechanism in the turbulent
fluid present after a phase transition \cite{H,McL}.

In this section, it was argued that although
magnetic fields are non-equilibrium 
configurations that break $C$, $CP$ and
$SO(3)$, the only ingredient required
to generate them is some out-of-thermal-equilibrium
process. This is in contrast to the
baryon asymmetry, where the symmetries that
are broken by the presence of the asymmetry have to be broken
to generate it.

\section{Maxwell's equations in the presence of a varying Higgs vev}

During the electroweak phase transition, 
the Higgs vev is in the process of
going from an SU(2) $\times$ U(1) symmetric ground state
to the vacuum state we live
in today. Vachaspati \cite{V} has argued that it
creates magnetic fields in this process.
Electromagnetism is difficult to
understand during the phase transition because the Higgs field, whose
vev usally defines the direction of the unbroken symmetry, is space-time
and gauge dependent. To understand the implications
of the space-time variations in the Higgs, one
must separate them
from the gauge dependence.

  't Hooft has given a gauge-invariant
definition of
the electromagnetic field $F_{\mu \nu}$ in the Georgi-Glashow
model (where  SO(3) is broken  to U(1)), in the
presence of a non-trivial Higgs vev.
Vachaspati \cite{V} similarly defined  a gauge-invariant 
 electromagnetic tensor for the  Standard Model
\be
F_{\mu \nu}^{em} = \sin \theta_W \eta^a W^a_{\mu \nu} + 
\cos \theta_W B_{\mu \nu} - i\frac{4 \sin \theta_W}{g \phi^{\dagger} \phi}
[({\cal D}_{\mu} \phi)^{\dagger} {\cal D}_{\nu} \phi
- ({\cal D}_{\nu} \phi)^{\dagger} {\cal D}_{\mu} \phi] \label{VF}
\ee
where 
\be
{\cal D}^{\mu} = \partial^{\mu} - i g' \frac{Y}{2} B^{\mu}
- i g \vec{T} \cdot \vec{W}^{\mu} \label{cd}
\ee
 ($T_i = \sigma_i/2$).
He then argues that (\ref{VF}) is naturally non-zero
during the phase transition, because the 
varying Higgs vevs are uncorrelated across the horizon. 
It is unclear (to this author) whether such a complicated
expression is zero or not, although it does have simpler,
but less obviously gauge invariant formulations. The
purpose of writing
this expression was to have a gauge
invariant formulation, so  arguing
that the $B_{\mu \nu}$ and $ W_{\mu \nu}$ part
of this is zero, and the scalar part
is not is a gauge dependent statement. It would
be clearer to retain the explicit gauge
independence.
Let us  therefore assume that Maxwell's equations
are correct, so that  one cannot generate
a field without a source. 
In a purely classical analysis, 
the electromagnetic field evolves as
\be
\partial_{\mu} F^{\mu \nu} = j^{\nu} \label{j}
\ee
where $j^{\nu}$ is the current due the
the rolling classical Higgs vev (and not
the expectation value of the quantum operator $j^{\nu}$). 
Therefore if 
$F_{\mu \nu} = 0$ at the beginning of the phase
transition, it is zero at the end unless there is some source
during the transition. 

The point is now to show that one
 can write an SU(2)$\times$U(1) gauge invariant
definition of $j^{\nu}$ that is clearly zero
during the phase transition. 

  In ordinary QED, 
for a singly charged
scalar, the RHS of equation (\ref{j}) is
\be
j^{\nu} = i e \phi^* Q {\cal D}^{\nu} \phi
- i e ({\cal D}^{\nu}\phi)^* Q  \phi \label{j2}
\ee
where ${\cal D}^{\nu}  = \partial^{\nu} - i e Q A^{\nu}$,
and $Q$ is the charge operator.
This can  be generalized  to SU(2) $\times$ U(1),
following 't Hooft;  if one varies the Standard Model  Higgs kinetic
term with respect to  the photon field $A^{\nu}$, one gets
\be
\frac{\delta}{\delta A^{\nu}} ({\cal D}^{\mu} \phi)^{\dagger} 
({\cal D}_{\mu} \phi) =  i e \phi^{\dagger} Q {\cal D}^{\nu} \phi
- i e ( {\cal D}^{\nu}\phi)^{\dagger} Q  \phi
\ee
where ${\cal D}_{\mu}$ is defined in (\ref{cd}), and
$Q$ is the gauge invariant generator in the
direction of the unbroken symmetry, which is  defined
 in equation (\ref{Q}).
Under an SU(2) $\times$ U(1)
gauge transformation $U$,
one can see  that
\be
j^{\nu}   \longrightarrow i e \phi^{\dagger}U^{\dagger} Q 
U{\cal D}^{\nu} \phi -  i e ( {\cal D}^{\nu}\phi)^{\dagger}U^{\dagger} Q 
U \phi
\ee
so   $j^{\nu}$  is gauge invariant if $Q= U^{\dagger} Q U$.  
One can check that
\be
Q = -2 \frac{ \phi^{\dagger} T^a \phi}{ \phi^{\dagger}  \phi} 
T^a - \frac{Y}{2} \label{Q}
\ee
is  gauge invariant under
both hypercharge and SU(2) transformations, and 
reduces to the usual definition of $Q$ in
the unitary gauge.

It is easy to check  that the operator $Q$ from
equation (\ref{Q}) is the  generator of the unbroken symmetry,
ie
\be
Q \phi = 0
\ee
so that there is no electromagnetic current,
and therefore no magnetic field, generated by the rolling
Higgs vev during the EPT.

A possible caveat to this argument is that
it only applies when $Q \neq 0$. However,
 the Standard Model has no topological
defects, so one can imagine deforming the Higgs
slightly away from zero,
and then this argument should
be applicable. 
 There are non-topological defects and other
classical field
configurations that have associated magnetic fields,
but the coherence length is presumably much shorter
than the horizon size, so that the
random walk average on galaxy scales \cite{H}
produces too small a field today.

\section*{conclusion}
It has been argued that the classical rolling Higgs  at the
electroweak phase transition does not generate a 
horizon-scale magnetic field.
However, it is not excluded
that thermal or quantum fluctuations about
this time dependent vev 
could produce the $\vec{B}$ field.

The symmetries and
conditions broken by the presence of a magnetic field, namely
thermal equilibrium, $CP$, $C$, and
rotational invariance have  been identified.
It is possible for the magnetic field to
spontaneously break $C$, $CP$ and $SO(3)$ as it
develops, so these symmetries do not
need to be broken to generate  $\vec{B}$. 
However, some out-of-thermal-equilibrium physics
is required. 

\subsection*{acknowledgements}
I would like to thank    Tim Evans
for many discussions, Tanmay Vachaspati  for 
clarifying reference \cite{V},  and Georg Raffelt for helpful
comments. I am also grateful to  
 Steve Arendt, Luis Bettencourt,  Paolo Gambino, David Jackson,
Tom Kibble, Ray Rivers, Joe Silk  and Sharon Vadas for
useful conversations.

\end{document}